\documentclass[usenatbib]{mn2e}
\usepackage{psfig,times}

\title[Spurious `active longitudes']{Spurious `active longitudes' in parametric 
models of heavily spotted eclipsing binaries}

\author[S.V. Jeffers]
       {S.V. Jeffers$^1$ \\ $^1$School of Physics and Astronomy,
	University of St Andrews,
       North Haugh, St Andrews, Fife KY16 9SS, U.K.}

\date{}

\pagerange{\pageref{firstpage}--\pageref{lastpage}}
\pubyear{2005}

\begin{document}

\maketitle

\label{firstpage}

\begin{abstract}
In this paper, size distributions of starspots extrapolated from the
case of the Sun, are modelled on the eclipsing binary SV Cam to
synthesise images of stellar photospheres with high spot filling
factors.  These spot distributions pepper the primary's surface with
spots, many of which are below the resolution capabilities of eclipse
mapping and Doppler imaging techniques.  The lightcurves resulting
from these modelled distributions are used to determine the
limitations of image reconstruction from photometric data.  Surface
brightness distributions reconstructed from these lightcurves show
distinctive spots on the primary star at its quadrature points.  It is
concluded that two-spot modelling or chi-squared minimisation
techniques are more susceptible to spurious structures being generated
by systematic errors, arising from incorrect assumptions about
photospheric surface brightness, than simple Fourier analysis of the
light-curves.
\end{abstract}
\begin{keywords}

stars: activity, spots, imaging  binaries: eclipsing

\end{keywords}

\section{Introduction}
Doppler imaging has shown that rapidly rotating RS CVn binary systems,
such as SV Cam (F9\,V\,+\,K4\,V, P$_{rot}$=0.59\,d) frequently show
spots at high and polar latitudes.  A successful theoretical model of the
formation of polar spots assumes that magnetic flux from decaying
active regions is swept toward the poles by meridional flows
\citep{schrijver01polar}.  However, in order to produce 
polar spots bipolar active regions have to emerge at a rate
approximately 30 times faster than in the case of the Sun, implying
that the photospheres of active stars should be peppered with a large
number of small spots.

Jeffers et al.(2005) used spectrophotometric data from the {\em Space
Telescope Imaging Spectrograph} on board the Hubble Space Telescope to
eclipse map the inner face of the primary of the RS CVn SV Cam.  These
observations and the {\sc hipparcos} parallax showed that the surface
flux in the eclipsed low latitude region is approximately 30\% lower
than computed from the best fitting {\sc phoenix} \citet{allard00}
model atmosphere.  This flux deficit can only be accounted for if
approximately 28\% of the primary's surface is peppered with dark
spots too small to be resolved through eclipse-mapping techniques
where the resolution limit is approximately 1.5 degrees.

\begin{table*}   
\label{input}
%\fontsize{10}{14}\selectfont 
\begin{tabular}{l c c c c c c l }
\hline
\hline
Set & 1 & 2 & 3 & 4 & 5 & 6 & \,\,\,7 \\
\hline
Spot Coverage & &  0.3\% & 1.6\% & 6.1\% & 18\% & 48\% & 100\% \\
{$\sigma_A$}& 3.8 & 5.0 & 6.8 & 9.2 & 12.2 & 15.8 & 20.0\\
{(dN/dA)$_{max}$} & 5 & 25 & 65 & 125 & 205 & 305 & 425 \\
{A$_{spot,tot}$} ($\times$10$^{-6}$) & 320 & 3000 & 16000 & 61000 & 186000 & 484000 &
10$^6$\\
{A$_{max}$} (A$_{*}$) & 200 & 400 & 1000 & 3000 & 6000 & 10000 &
10000 \\ 
\hline 
{Solar:}&{min}&{max}\\
\hline
\hline 

\end{tabular}
\caption{Tabulation of the input parameters to the log-normal size
distribution of star spots, equation~\ref{equ1}.  The parameters
derived by \citet{bogdan88}
for the Sun are data sets 1\&2, and those calculated by \citet{solanki99}
for active stars are data sets 3-7. }     
\end{table*}   

This paper extends the work of Jeffers et al. (2005) by applying the
extrapolated solar spot size distribution of \citet{solanki99} to a
hypothetically immaculate primary star of SV Cam.  We determine how
different degrees of spot coverage influence the shape of the binary
system's lightcurve, and how accurately these lightcurves are
reconstructed into surface brightness distributions using the Maximum
Entropy (Max Ent) eclipse mapping technique (\citealt{cameron97dots};
\citealt{cameron97xyuma}).

\section{Modelled Data}

\subsection{Size distribution of spots on active stars}

The variable nature of the spot coverage of active stars makes the
quantification of their spot size distribution an intriguing problem.
In the case of the Sun, the spot size distribution has been determined
by \citet{bogdan88} from direct observations taken from the Mount
Wilson white-light plate collection covering the period 1917-1982.
When plotted on a log-log scale the size distribution of sunspots is
parabolic, implying a two-parameter log-normal distribution, which
cannot be represented by a single parameter distribution such as a
power law.  Following \citet{bogdan88} the number of sunspots, N, as a
function of the solar surface area, A, is equated as,

\begin{equation}  
\frac{dN}{dA} = \left(\frac{dN}{dA}\right)_{max}
\exp\left(-\frac{(\ln{A}-\ln{\langle A \rangle})^2}{2\ln{\sigma}_A}\right)
\label{equ1}
\end{equation}

\noindent where the constants $\langle$A$\rangle$ and $\sigma_A$ are the mean
and geometric standard deviation of the log-normal distribution, and
$\left(\frac{dN}{dA}\right)_{max}$ the maximum value reached by the
distribution.  For the case of the Sun, these values are tabulated in
Table~\ref{input} where set 1 is for an inactive Sun and set 2 is for
an active Sun.

The extrapolation of the solar spot size distribution to active stars
can be represented by different extrapolations e.g. \citet{solanki99}
and \citet{solanki04}.  As long as the size distribution shows many
sub-resolution spot that are not clumped together, the exact
representation of the extrapolated size distribution is of minor
importance.  The extrapolation of \citet{solanki99} is used, which is
determined by analysing how the observable parameters of the log
normal distribution, (dN/dA)$_{max}$, $\langle$A$\rangle$, and
$\sigma_A$, change with increasing magnetic activity.  The
extrapolated size distributions cover a range of degrees of
spottedness, from the one at solar minimum to a hypothetically
completely spotted star.  As adopted in the \citet{solanki99} model,
the penumbral diameter is included in the spot area (in the umbral to
penumbral area ratio of 1:3) and is assumed to be independent of spot
size.  In all cases the minimum spot area is
1.5$\times$10$^{-6}$A$_{\odot}$, where A$_{\odot}$ is the surface area
of the visible solar hemisphere.  The input parameters for the
log-normal distributions of \citet{bogdan88} and
\citet{solanki99} are tabulated in Table ~\ref{input}: sets 1
and 2 are for the inactive and active Sun, and sets 3 to 7 correspond
to active stars with a spot coverage of 1.8\%, 6.1\%, 18\%, 48\% and
100\% of their surface.  

\subsection{Generation of randomly distributed spots}

The binary eclipse mapping code DoTS \citep{cameron97dots} was used to
synthesise spot maps which follow a log-normal size distributions on
the surface of an immaculate SV Cam's primary (primary:
R$_1$=1.24R$_\odot$, T$_{eff}$=6038\,K, secondary:
R$_2$=0.79R$_\odot$, T$_{eff}$=4804\,K) to incorporate the surface
geometry and radial velocity variations of tidally distorted close
binary stars.  The input parameters to DoTS for modelling spots are, where
x is the output from the random number generator (0$\le$x$\le$1);
\newline
(i) {\bf longitude:} randomly distributed between
0$^\circ$ \& 360$^\circ$ \newline 
(ii) {\bf latitude:} $-\frac{\pi}{2}<
\theta<\frac{\pi}{2}$, following $\theta= \arcsin \left( 2x+1 \right)$ 
with 0$\le$x$\le$1, to eliminate an artificial concentration of spots
at the pole \newline (iii) {\bf spot radius:} computed using the
previously described log-normal distribution as tabulated in
Table~\ref{input} \newline (iv) {\bf spot brightness \& spot
sharpness:} modelled to obtain an umbral to penumbral ratio of 1:3.

\subsection{Spot distributions}

\begin{figure*}
\begin{center}
\begin{tabular}{ccc}
 \vspace{5mm} \\
 \psfig{figure=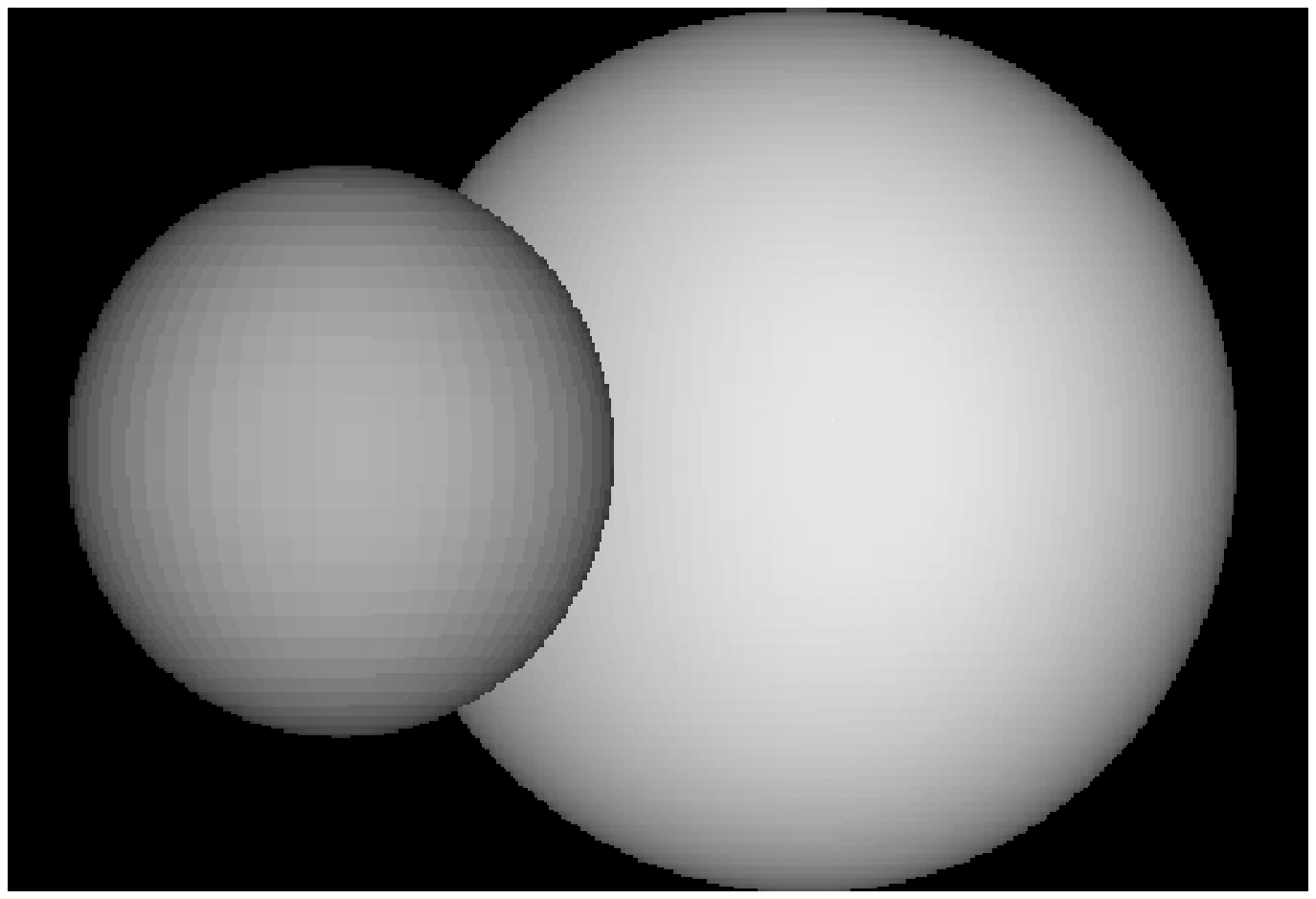,width=6cm,height=4cm,bbllx=15bp,bblly=15bp,bburx=405bp,bbury=280bp} &
\hspace{-4mm}
\hspace{1mm}
 \psfig{figure=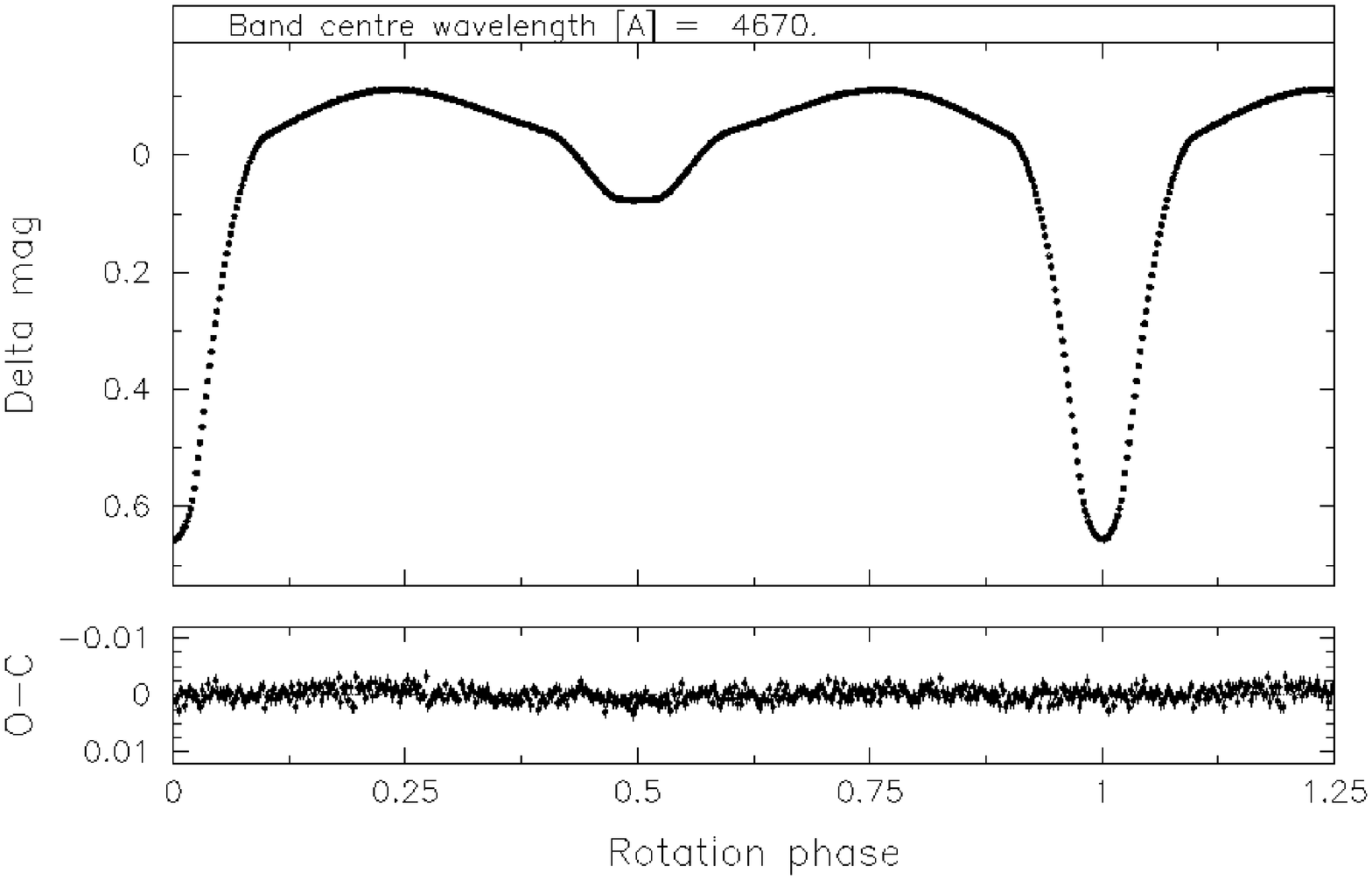,angle=270,width=6cm,height=4cm,bbllx=20bp,bblly=50bp,bburx=445bp,bbury=671bp} \\ \\ \\

 \psfig{figure=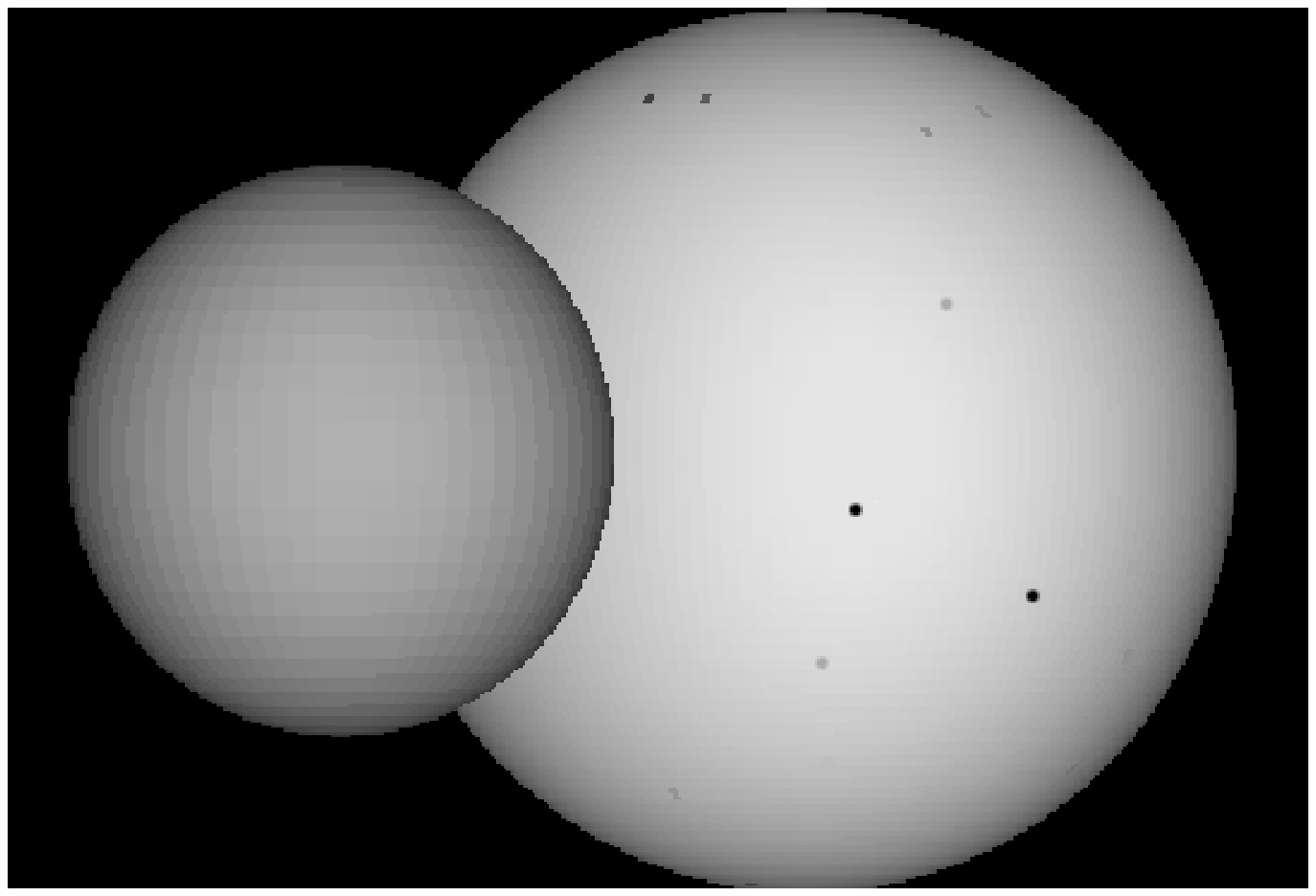,width=6cm,height=4cm,bbllx=15bp,bblly=15bp,bburx=405bp,bbury=280bp} &
\hspace{-4mm}
\hspace{1mm}
 \psfig{figure=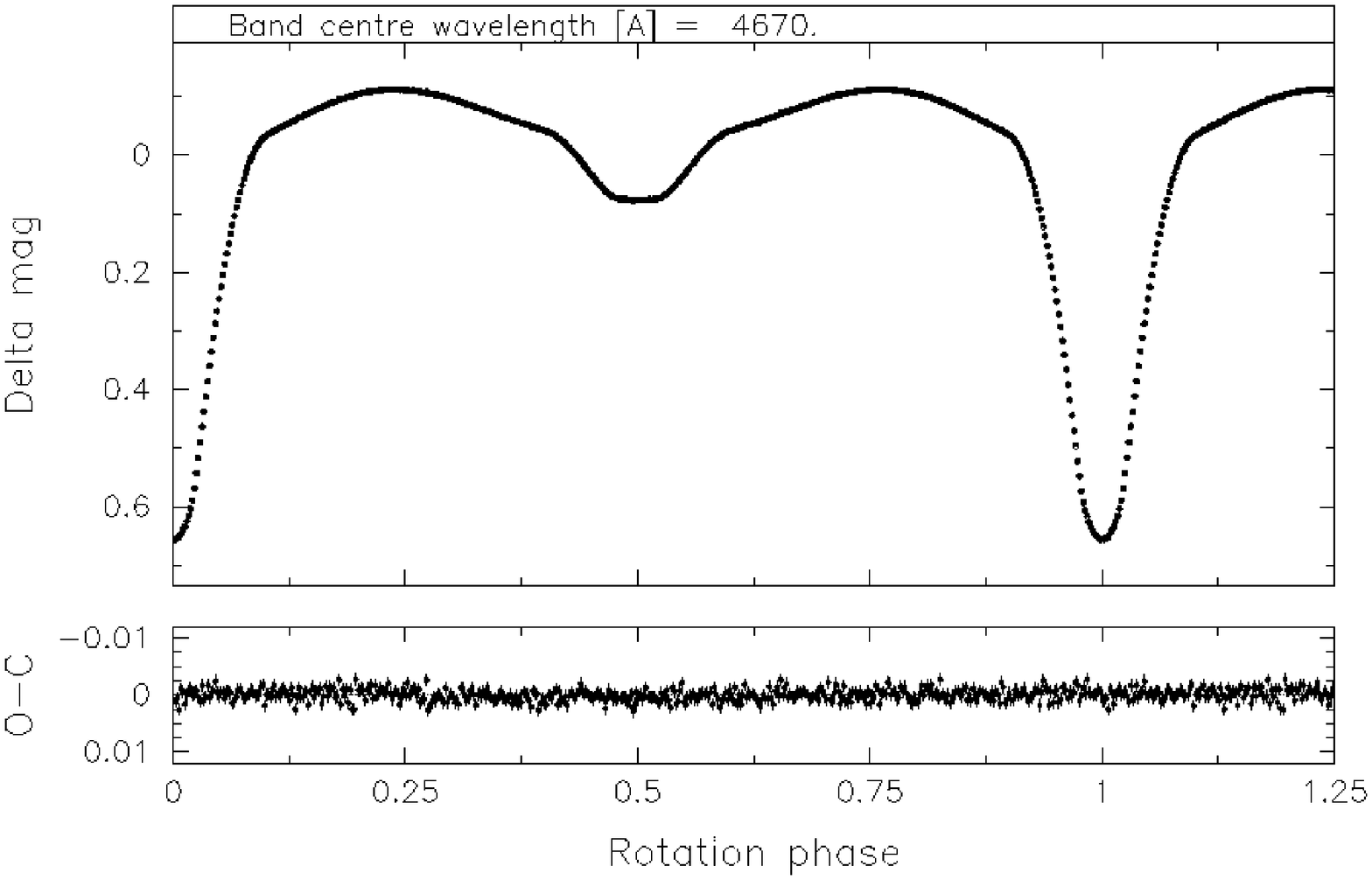,angle=270,width=6cm,height=4cm,bbllx=20bp,bblly=50bp,bburx=445bp,bbury=671bp} 
\hspace{1mm} \\ \\ \\

 \psfig{figure=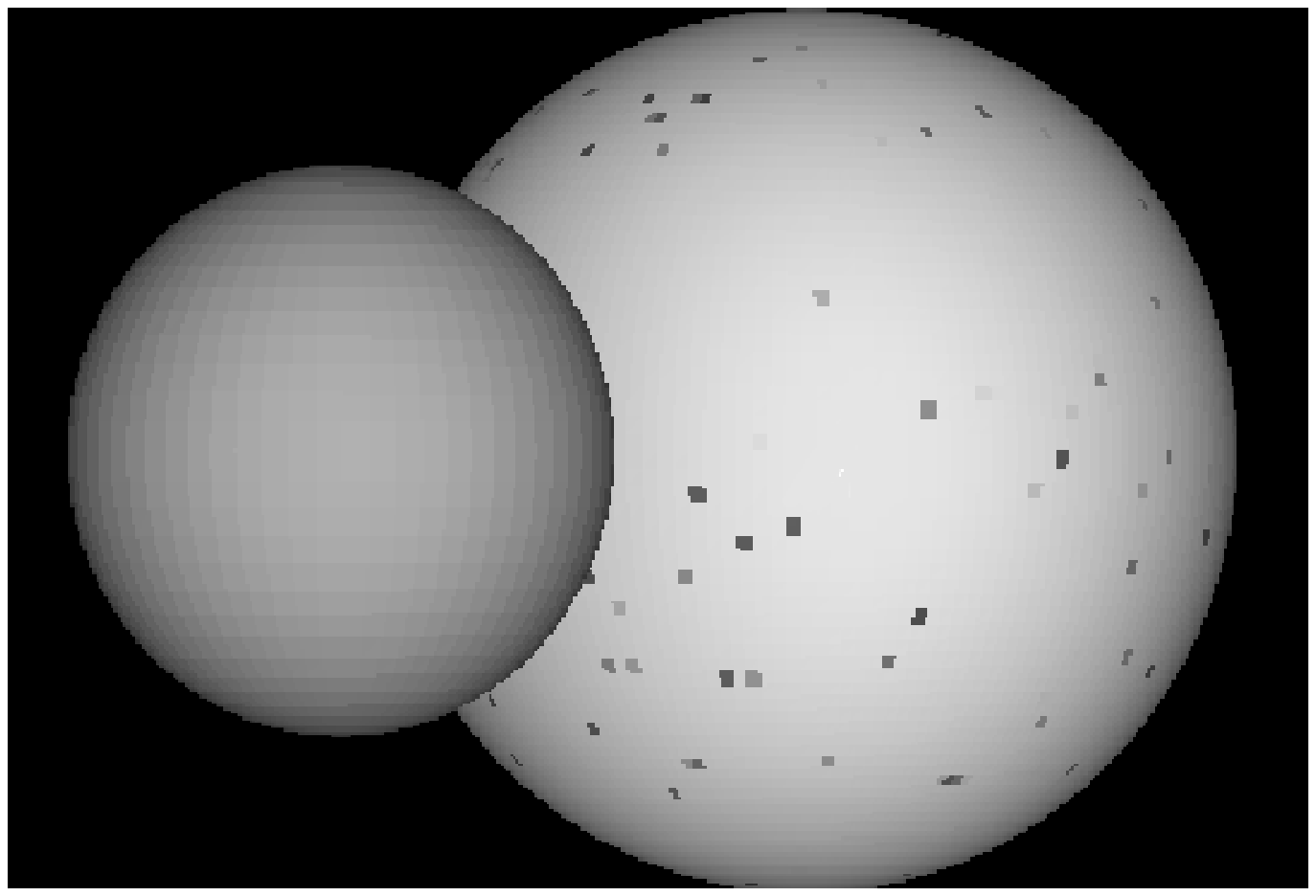,width=6cm,height=4cm,bbllx=15bp,bblly=15bp,bburx=405bp,bbury=280bp} &
\hspace{-4mm}
\hspace{1mm}
 \psfig{figure=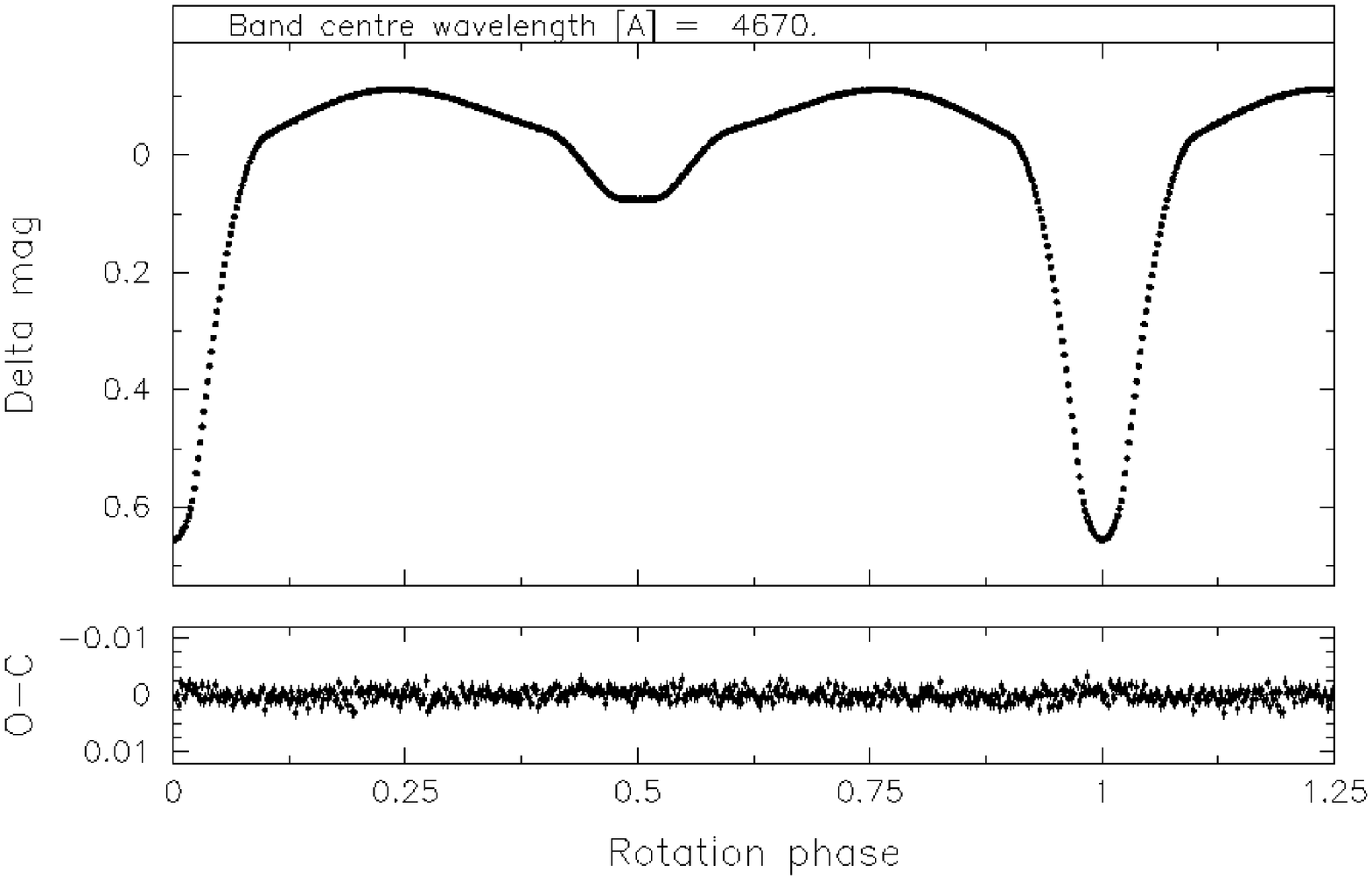,angle=270,width=6cm,height=4cm,bbllx=20bp,bblly=50bp,bburx=445bp,bbury=671bp} \\ \\ \\ 
\hspace{1mm}

\end{tabular}
\end{center}   
\vspace{5mm} 
\caption{(left) The distribution of spot size distributions for sets 1
to 3.  (right) The photometric lightcurve resulting from these spot
distributions and the Max Ent model fit to these lightcurves.  The
Observed minus Computed lightcurves are plotted at the bottom of each
lightcurve.  It was not possible to reconstruct surface brightness
distributions for these three sets as the spot coverage is too low to
be recovered using eclipse-mapping.}
\label{set1t3}
\end{figure*}

\begin{figure*}
\begin{center}
\begin{tabular}{ccc}
 \vspace{5mm} \\
\hspace{0cm}

\psfig{figure=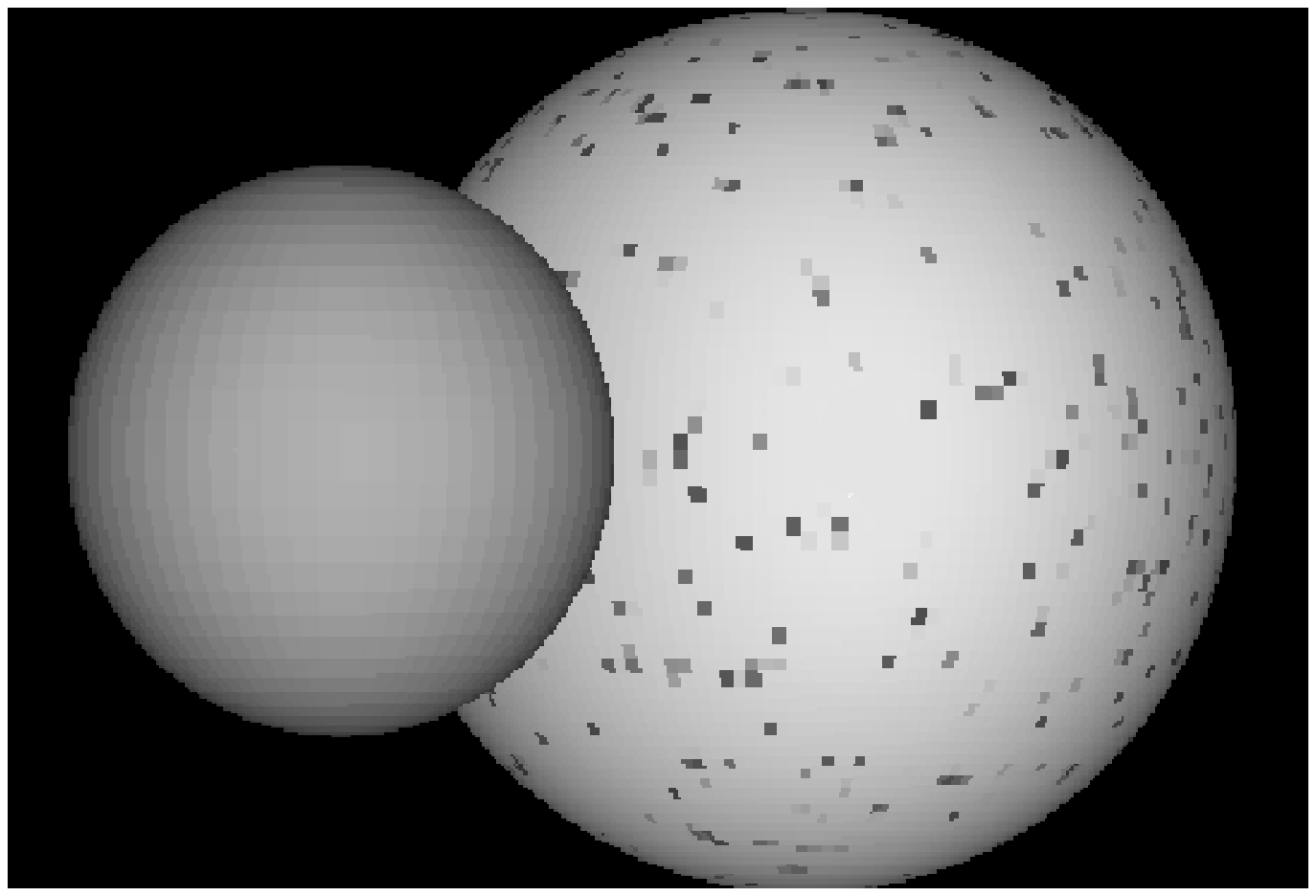,width=6cm,height=4cm,bbllx=15bp,bblly=15bp,bburx=405bp,bbury=280bp} &
\hspace{-3mm}
\hspace{-2mm}
 \psfig{figure=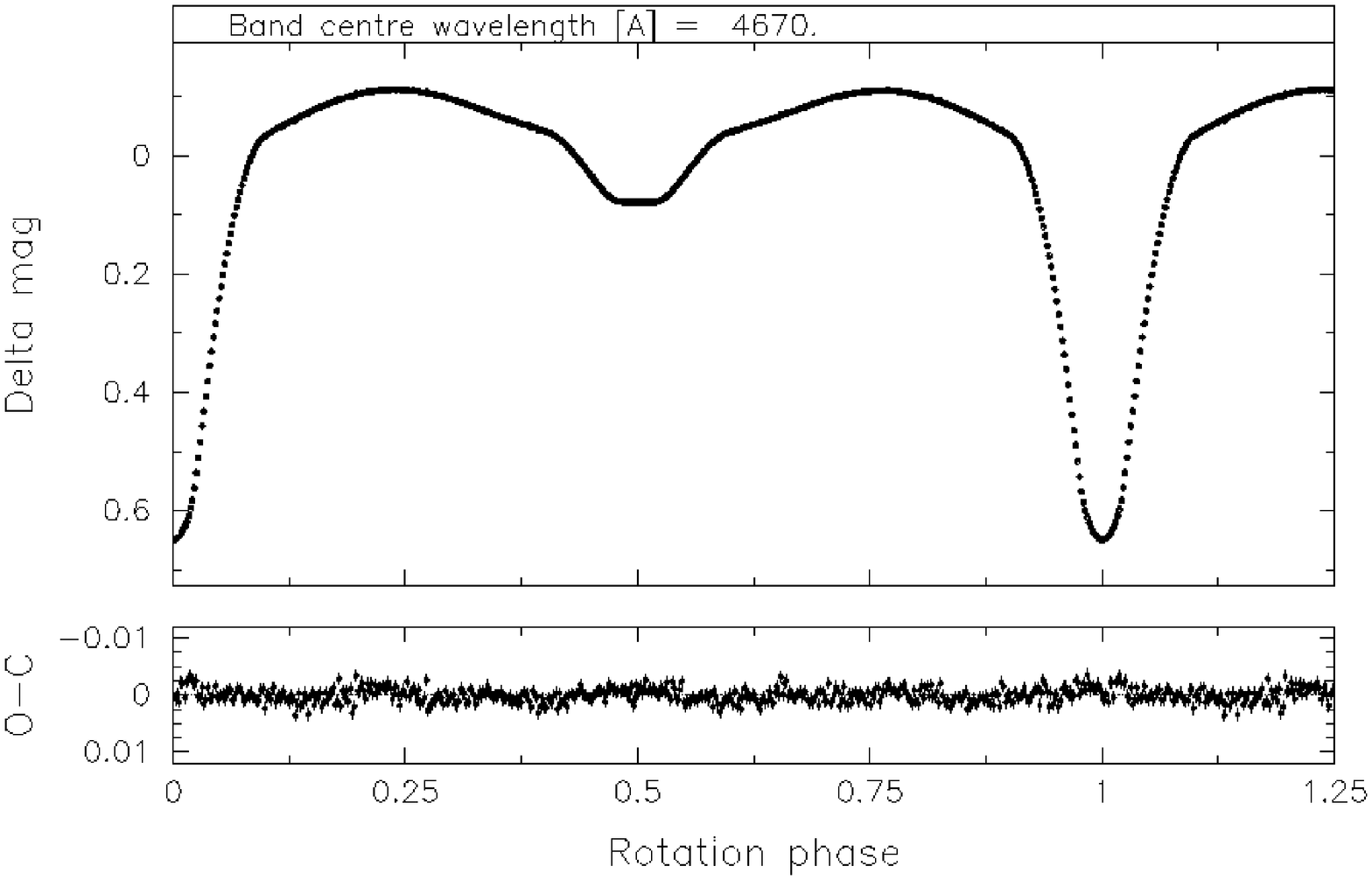,angle=270,width=6cm,height=4cm,bbllx=20bp,bblly=50bp,bburx=445bp,bbury=671bp}
\hspace{1mm}
 \psfig{figure=fig9.ps,angle=270,width=6cm,height=4.15cm,bbllx=355bp,bblly=12bp,bburx=60bp,bbury=601bp} \\ \\ \\

\psfig{figure=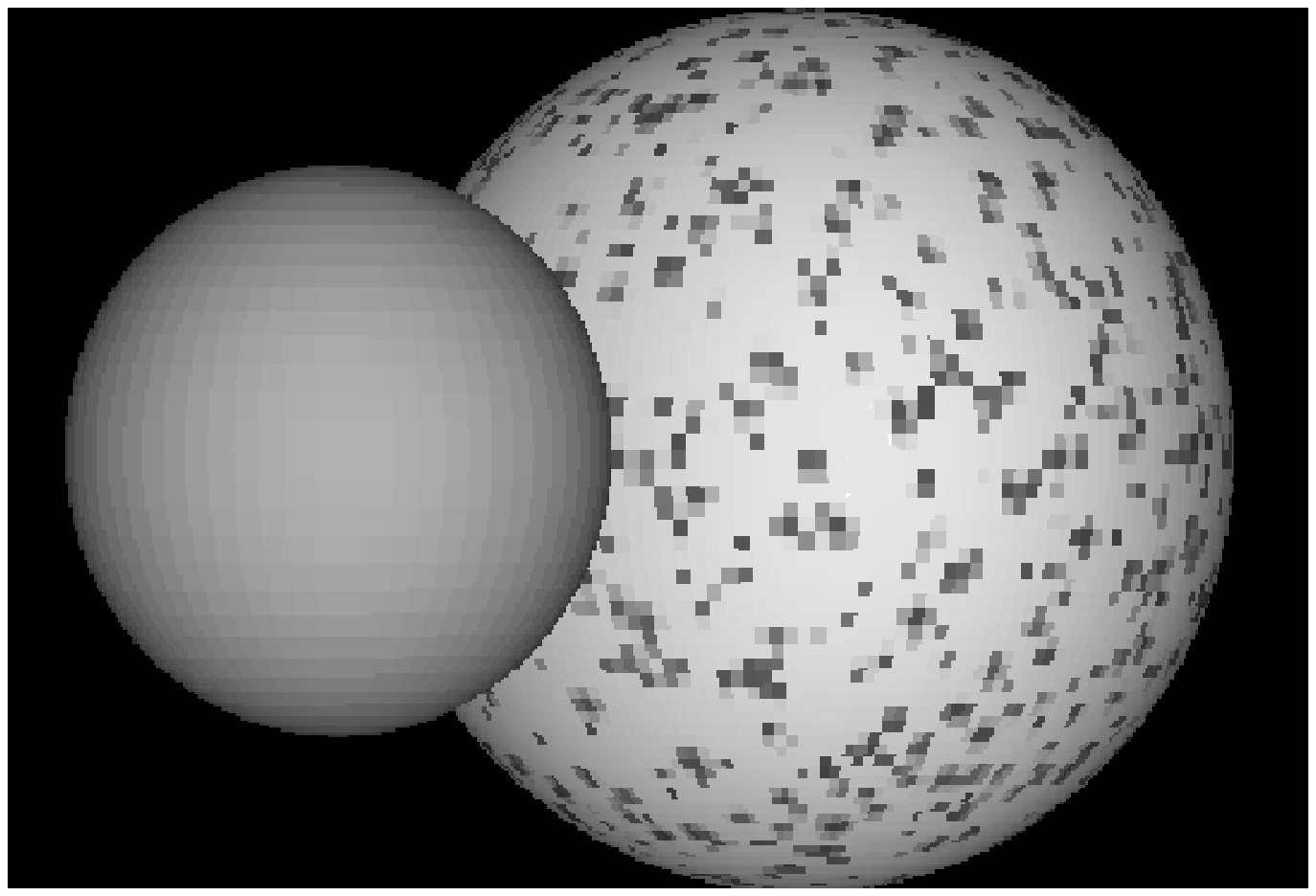,width=6cm,height=4cm,bbllx=15bp,bblly=15bp,bburx=405bp,bbury=280bp} &
\hspace{-3mm}
\hspace{-2mm}
  \psfig{figure=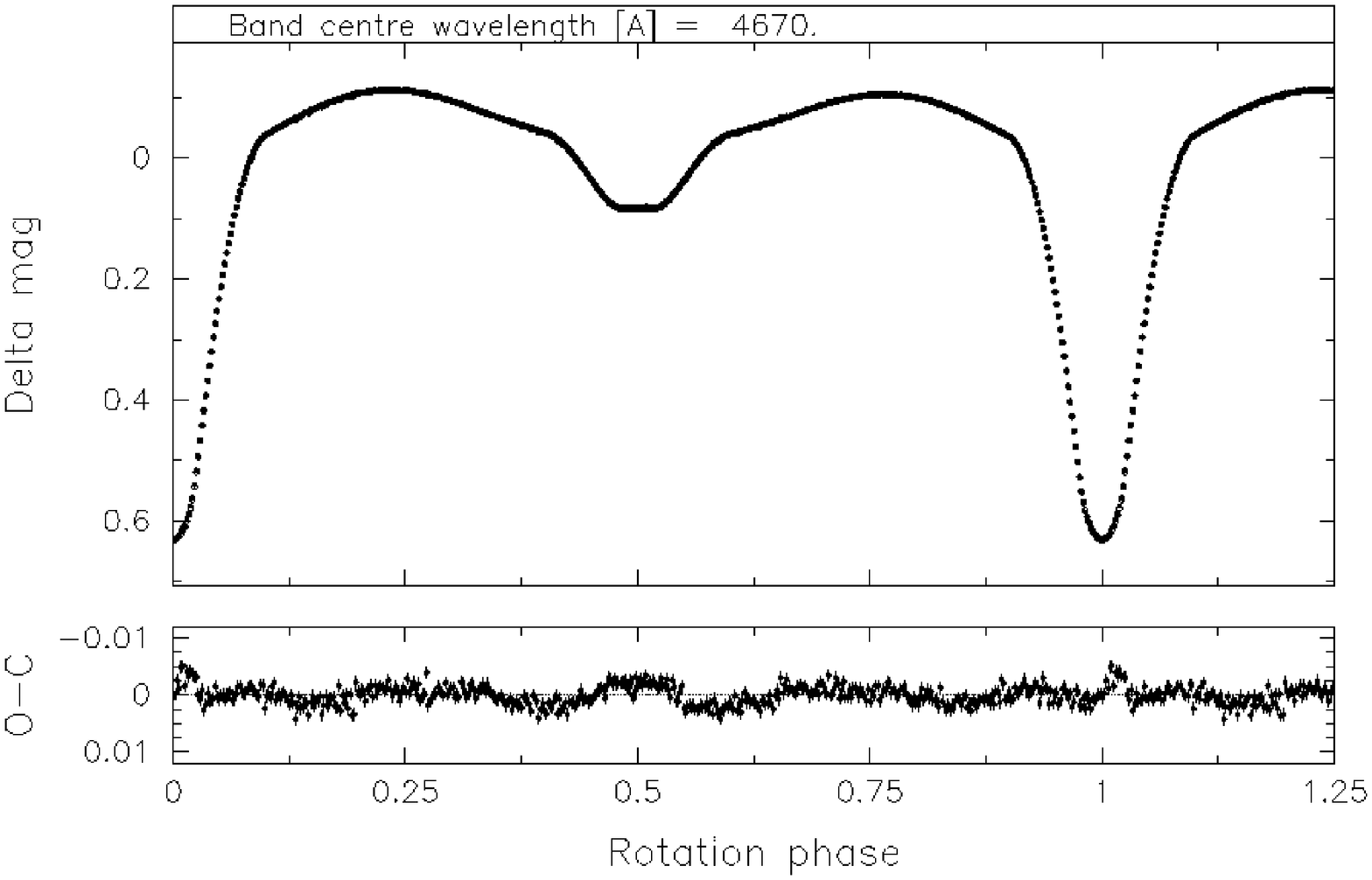,angle=270,width=6cm,height=4cm,bbllx=20bp,bblly=50bp,bburx=445bp,bbury=671bp}
\hspace{1mm}
 \psfig{figure=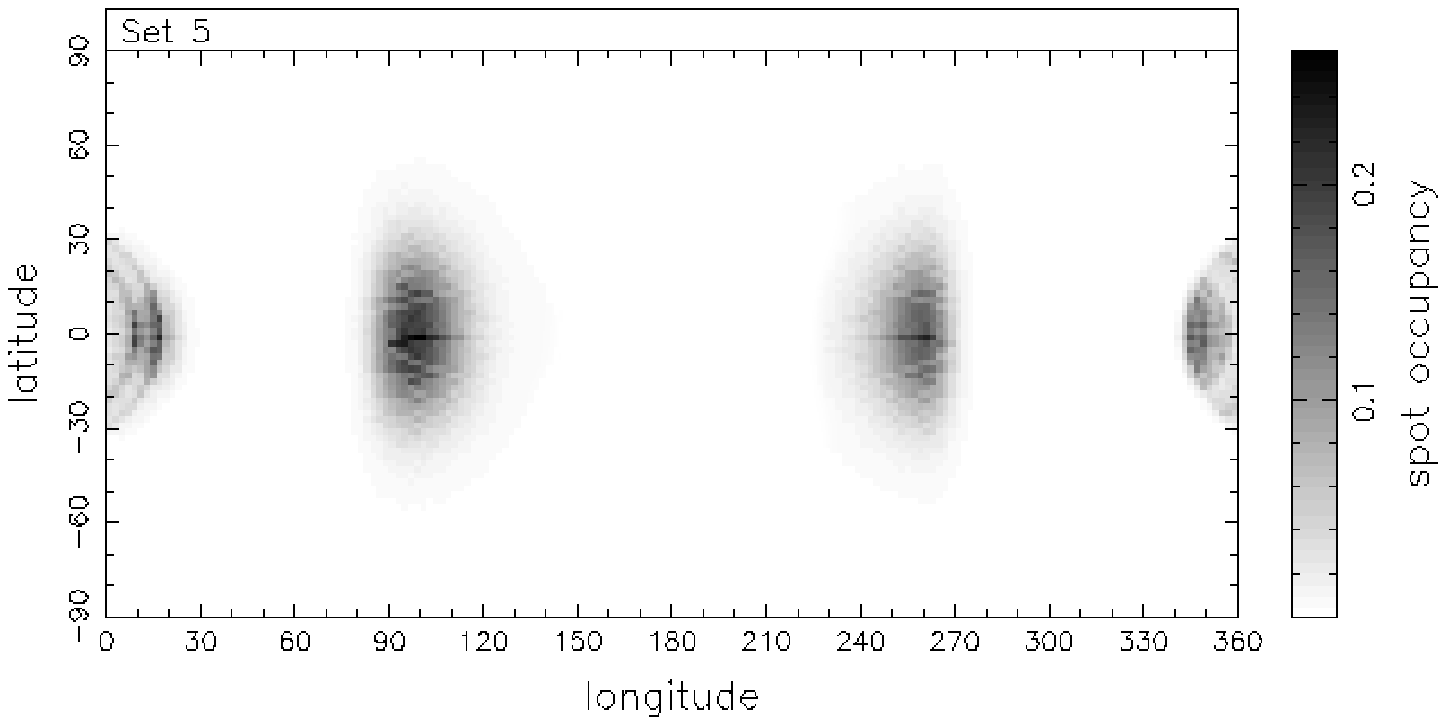,angle=270,width=6cm,height=4.15cm,bbllx=54bp,bblly=28bp,bburx=276bp,bbury=447bp} \\ \\ \\

\psfig{figure=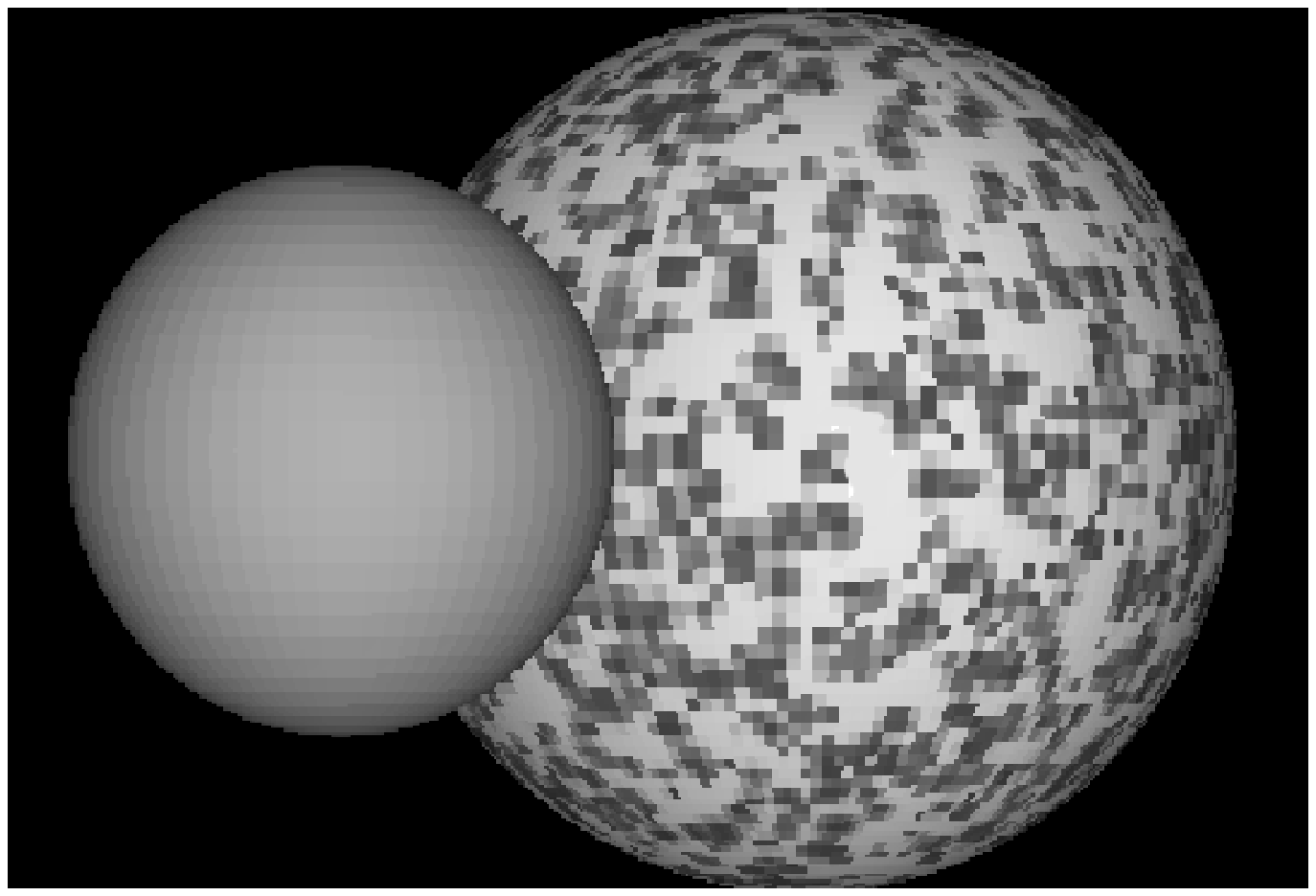,width=6cm,height=4cm,bbllx=15bp,bblly=15bp,bburx=405bp,bbury=280bp} &
\hspace{-3mm}
\hspace{-2mm}
 \psfig{figure=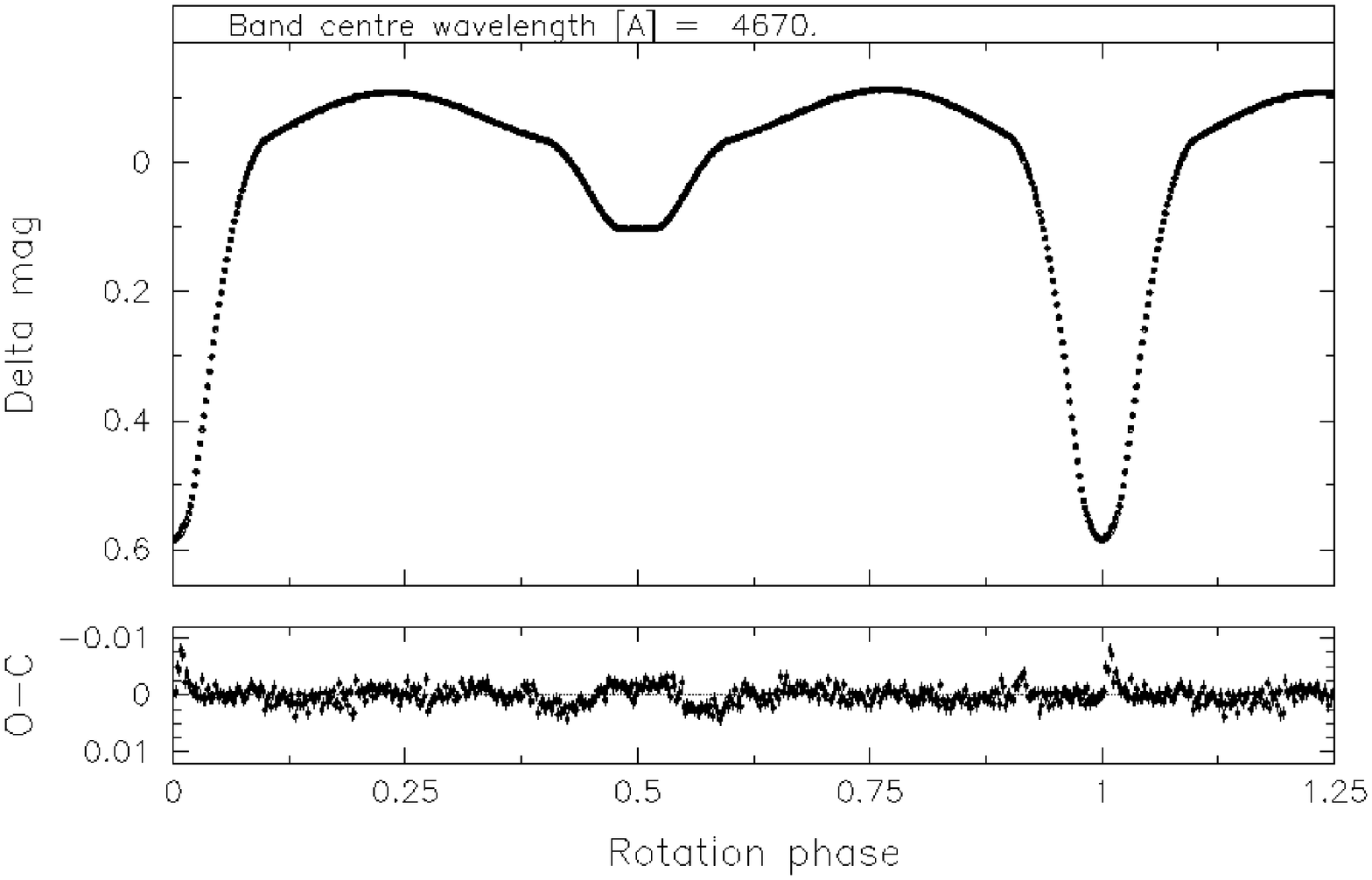,angle=270,width=6cm,height=4cm,bbllx=20bp,bblly=50bp,bburx=445bp,bbury=671bp}
\hspace{1mm}
 \psfig{figure=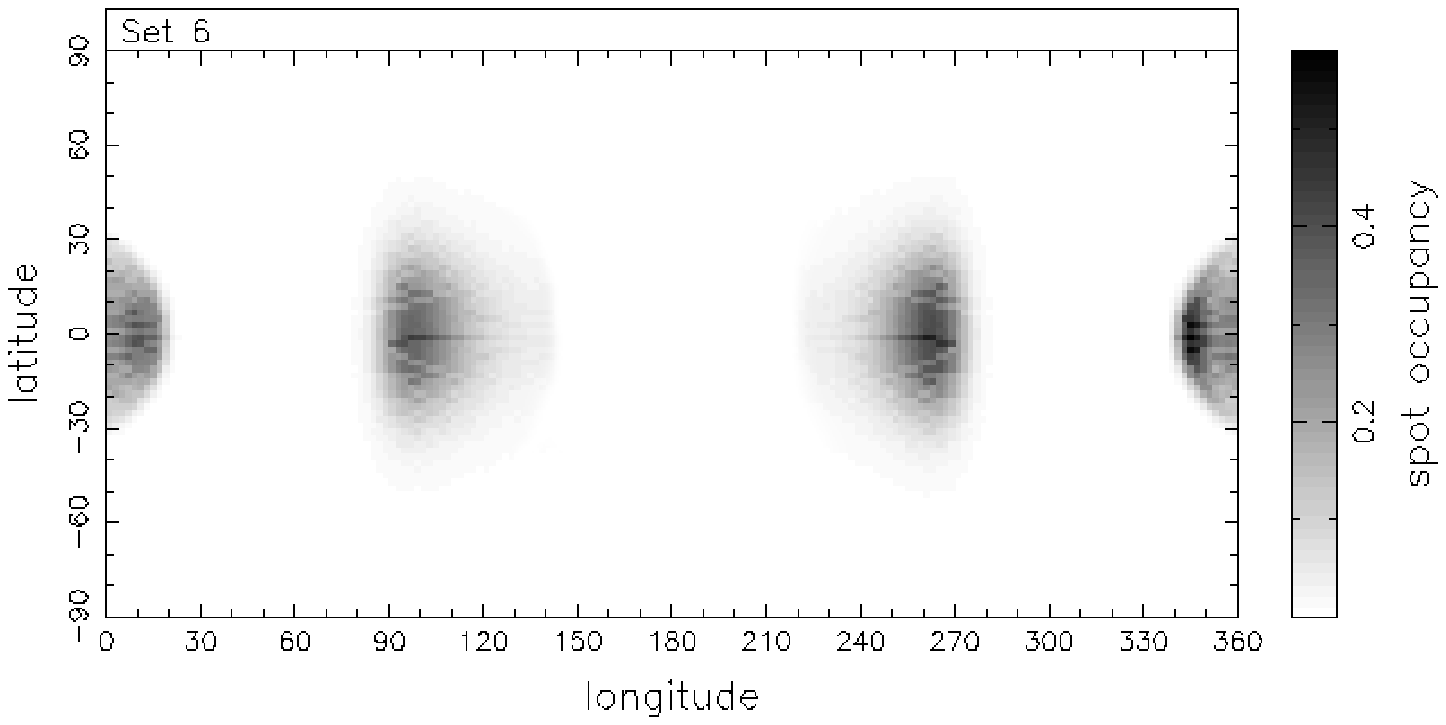,angle=270,width=6cm,height=4.15cm,bbllx=54bp,bblly=28bp,bburx=276bp,bbury=447bp} \\ \\ \\

\psfig{figure=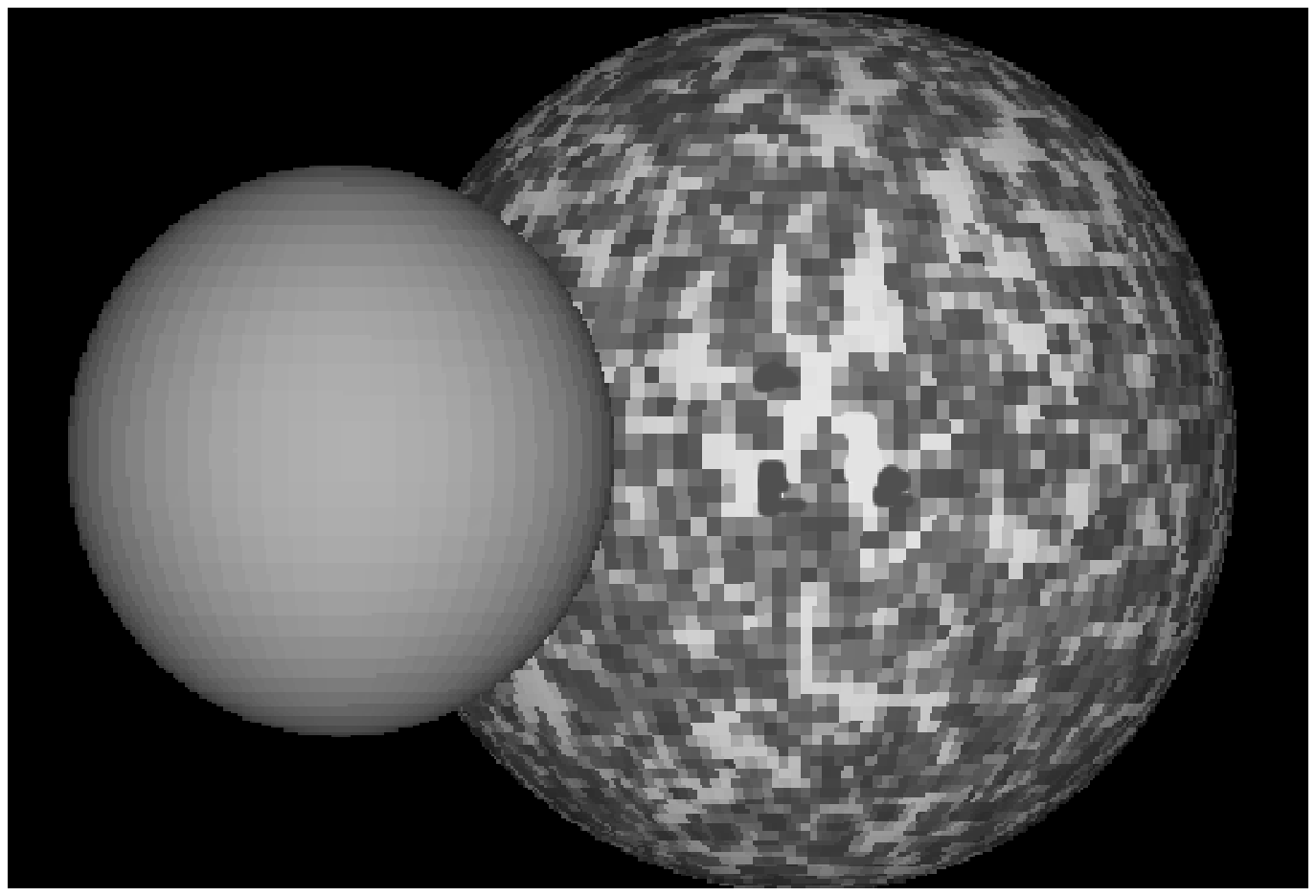,width=6cm,height=4cm,bbllx=15bp,bblly=15bp,bburx=405bp,bbury=280bp} &
\hspace{-3mm}
 \psfig{figure=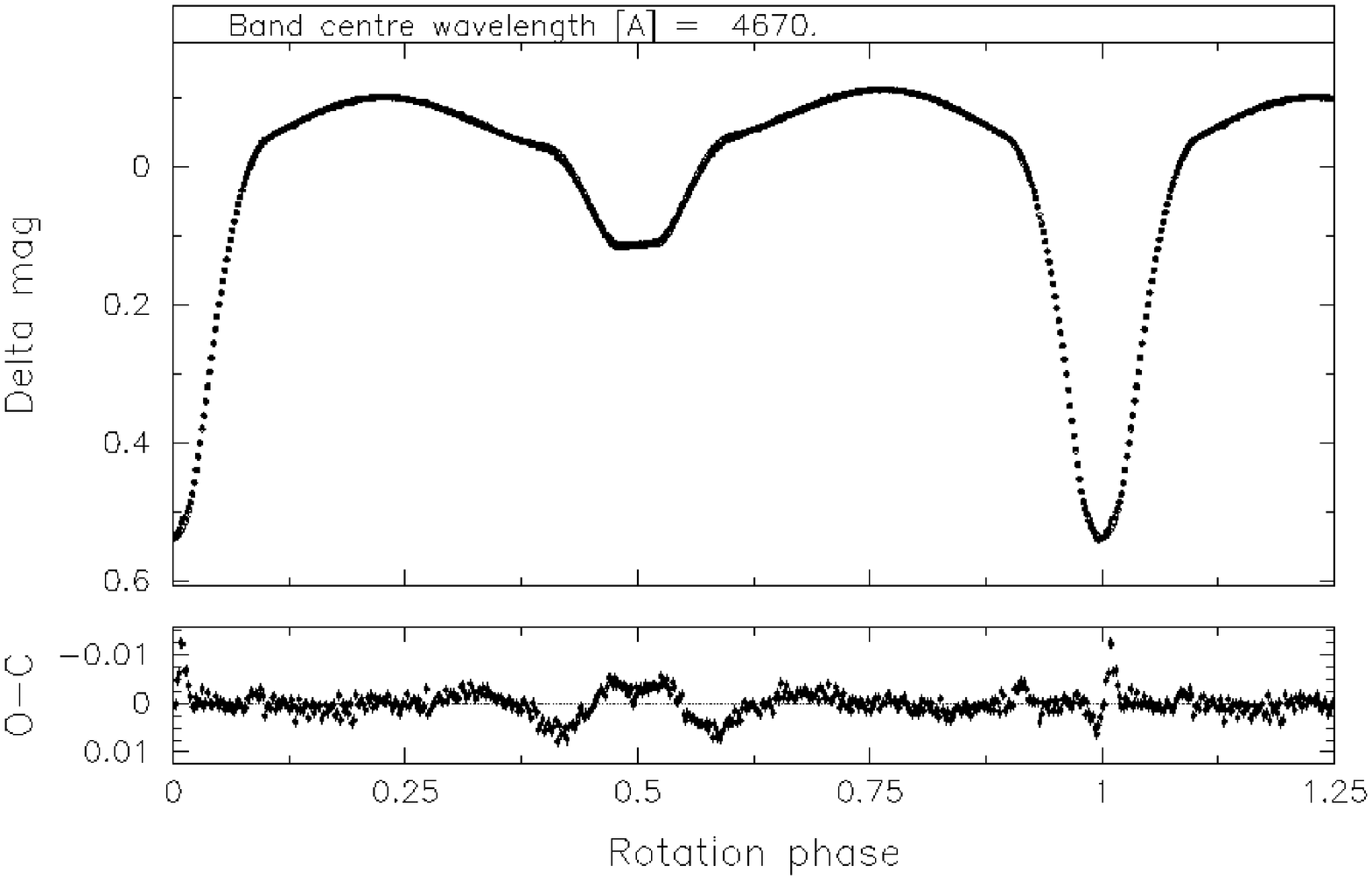,angle=270,width=6cm,height=4cm,bbllx=20bp,bblly=50bp,bburx=445bp,bbury=671bp}
\hspace{1mm}
 \psfig{figure=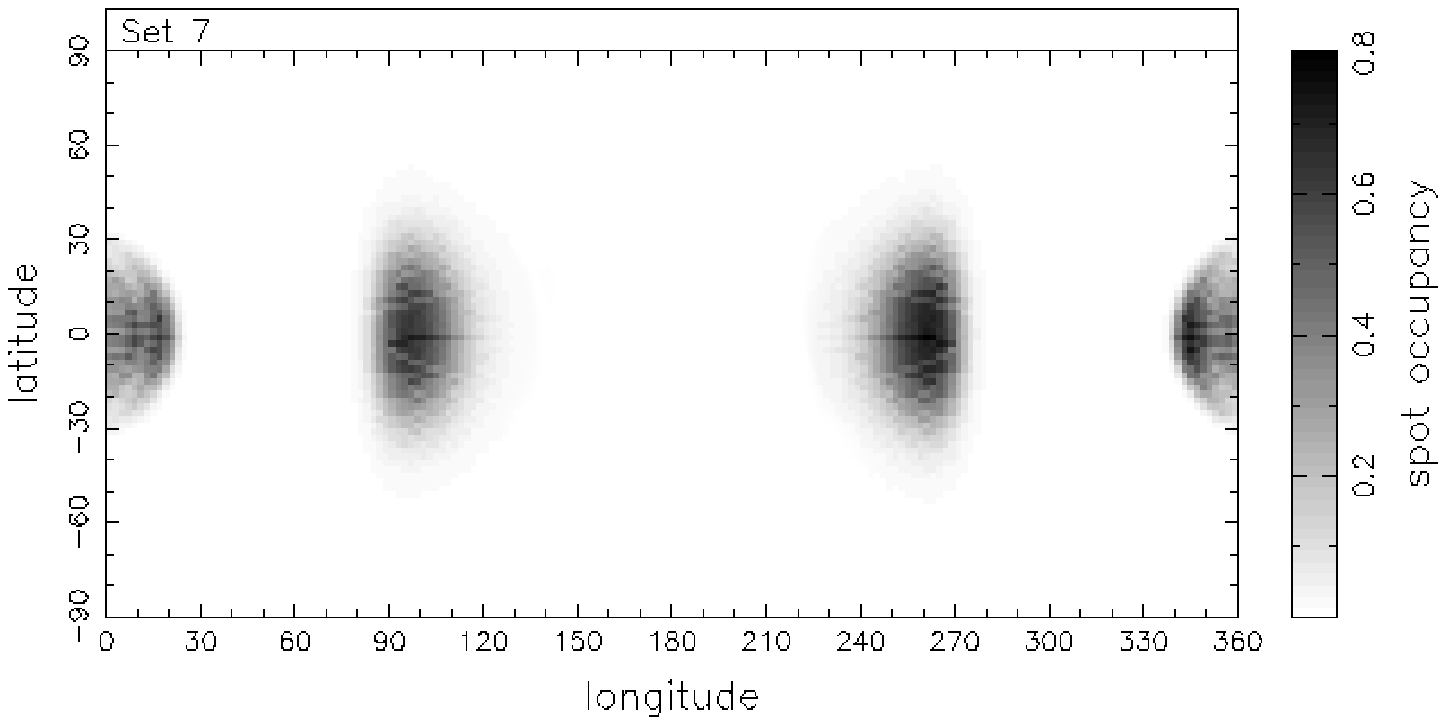,angle=270,width=6cm,height=4.15cm,bbllx=54bp,bblly=28bp,bburx=276bp,bbury=447bp}

\end{tabular}
\end{center}   
\vspace{5mm} 
\caption{(left) The modelled distribution of spots for sets 4 (top) 
to 7 (bottom).  (centre) The photometric lightcurve resulting from
these spot distributions and the Max Ent model fit to these
lightcurves.  The Observed minus Computed lightcurves are plotted at
the bottom of each lightcurve. (right) The surface brightness
distribution that are reconstructed using the Max Ent eclipse-mapping
technique, where phase runs in reverse to longitude.  Note the large
spurious spots at the quadrature points.}
\label{set4t7}
\end{figure*}

The spot distributions resulting from the size distributions given in
Table~\ref{input} are shown on the left hand side of
Figure~\ref{set1t3} (sets 1 to 3) and in Figure~\ref{set4t7} (sets 4
to 7).  The degree of spottedness increases from the set at solar
minimum (top of Figure~\ref{set1t3}) to the set with 100\% spot
coverage (bottom of Figure~\ref{set4t7}) to show the extreme effect of
spot coverage.  In this case, as the spot coverage fraction includes
the spot's umbra + penumbra, the result is not a completely black
star.

These modelled spot distributions were used to generate a synthetic
lightcurve, comprising 500 points with uniform sampling and with
random Gaussian noise (0.0004) added to match the photometric
precision of our HST data.  The central lightcurve wavelength,
4670\AA, was also chosen to match that of our HST data.

\section{Surface Brightness Image Reconstruction}

The resulting lightcurve from the modelled distributions of spots, as
described in the previous section, is used as input to DoTS to
reconstruct the surface brightness distribution using the maximum
entropy $\chi^2$ minimisation method \citep{cameron97dots}.

\subsection{System parameters}

In the reconstruction of a surface brightness distribution of a binary
star, it is imperative that the correct binary system parameters are
used \citep{vincent93}.  Incorrect binary system parameters lead to a
poor fit of the photometric lightcurve and to the appearance of
spurious spot features on the surface brightness image to compensate
for this.  In this work, the assumed spot temperature used is 1500\,K
cooler than the star's photospheric temperature.  As the spot coverage
increases, the surface brightness of the star will decrease compared
to the star's photospheric brightness.  This is particularly relevant
for sets 4 to 7, where there is a high degree of spot coverage.  In
reconstructing the surface brightness distributions for each of these
sets the photospheric temperature, the primary radius and the
secondary radius are treated as an unknown parameters.  The best
fitting parameters are determined using a `grid search' method that
uses, as input to DoTS, a grid of: photospheric temperatures ranging
from 5500\,K to 6100\,K at 25\,K intervals; primary radius values
ranging from 1.22 to 1.26 $R_\odot$ and secondary radius values
ranging from 0.77 to 0.81 $R_\odot$, both in 0.005 $R_\odot$
intervals.  For each set, the lowest $\chi^2$ value is determined by
the minimum of a quadratic fit to these values.  The resulting
best-fitting parameters are summarised in Table~\ref{obs2}.  These
values are used as the photospheric temperature and radii for the
reconstruction of the primary's surface brightness distribution.

\begin{table}

%\fontsize{6}{8}\selectfont

\begin{tabular}{c c c c }
%\begin{tabular}{ l l l l l l }

\hline
\hline

{Set} & {Primary Temperature (K)} & {R$_{pri}$/$R_\odot$} & {R$_{sec}$/$R_\odot$} \\

\hline

initial & 6038 & 1.25 & 0.81 \\
4 & 6026$\pm$ 40 & 1.230$\pm$0.003 & 0.792$\pm$0.003  \\
5 & 5948$\pm$ 71 & 1.248$\pm$0.002 & 0.804$\pm$0.003 \\
6 & 5849$\pm$ 53 & 1.251$\pm$0.002 & 0.782$\pm$0.004 \\
7 & 5760$\pm$ 45 & 1.252$\pm$0.003 & 0.773$\pm$0.005 \\ 

\hline
\hline
\end{tabular}
\caption{Best fitting binary system parameters for the size distribution 
sets 4 to 7 in Table 1}
\label{obs2}
\end{table}

\subsection{Final image reconstruction}

The images reconstructed for sets 4 to 7 are shown in
Figure~\ref{set4t7}, however, the over-all spot coverage for sets 1 to 3
in Figure~\ref{set1t3} was too small to be reconstructed.  The
photometric lightcurves for sets 1 to 3 are shown in the right-hand
side of Figure~\ref{set1t3}.  For sets 4 to 7, the fit to the
photometric lightcurve and reconstructed surface brightness images are
shown respectively in the centre and the right-hand side of
Figure~\ref{set4t7}.

What is distinctive about these plots are the spot features at the
quadrature points (90$^\circ$ and 270$^\circ$).  These spot features
are not artifacts of the maximum entropy reconstruction, but occur as
a consequence of the $\chi^2$ minimisation technique.  If a star is
peppered with small unresolvable spots, the effect on the star's
lightcurve will be to reduce the depth of the primary eclipse.  To fit
this with $\chi^2$ minimisation it is necessary to increase the level
of the whole computed lightcurve to fit the reduced primary eclipse
depth.  The difference in light at the quadrature points can only be
accounted for by placing large spurious spots at these longitudes.

For sets 4 and 5, the spot feature at 270$^\circ$ longitude is fainter
than the spot feature at 90$^\circ$ longitude as there is a higher
degree of spot coverage at the first quadrature than at the second.
In the reconstructed surface brightness distributions, the spots at
0$^\circ$ longitude result from the total spot coverage between
second and third contact, i.e. during primary eclipse, while the spots
at 180$^\circ$ longitude result from spots on the primary star
visible during secondary eclipse.

\section{Discussion}

Size distributions of starspots extrapolated from the case of the Sun
\citep{solanki99}, have been modelled on the eclipsing binary SV Cam
to synthesise images of stellar photospheres with high spot filling
factors.  The lightcurves resulting from the extrapolated spot
distributions were used as input to the Max Ent eclipse-mapping code
DoTS.  The resulting surface brightness distributions, with
distinctive spot features at the quadrature points (90$^\circ$ and
270$^\circ$), bear little resemblance to the input distribution of
spots.  These spurious spots at the quadrature points
(90$^\circ$ and 270$^\circ$) are not artifacts of the eclipse mapping
technique, but a consequence of the $\chi^2$ minimisation method.

The presence of spots over a large fraction of the stellar surface
influences the shape of the lightcurve and consequently the derived
stellar parameters.  The peppering of the stellar surface with small
unresolvable spots decreases the depth of the primary eclipse,
resulting in a lower photospheric temperature of the primary star and
a corresponding change in the radii of both binary components.  The
change in the binary system parameters (cf Table 2) shows how the
degree of spottedness affects the determination of the correct
photospheric temperature and radii when only photometric observations
are used.

Spot features occurring at the quadrature points are commonly
reconstructed features of RS CVn binaries and typically referred to as
`active longitudes'.  They are interpreted as being the preferred
longitude of emergence of magnetic flux.  These features have been
observed on numerous short-period active RS CVn binary stars such as
EI Eri, II Peg, $\sigma$\,Gem, HR\,7275, AR Lac, SZ Psc, HK Lac
(\citealt{berdyugina98longitudes}; \citealt{lanza98};
\citealt{lanza01}; \citealt{jetsu96}; \citealt{henry95diffrot}; 
\citealt{olah91longitudes}).  In these cases, however, the `active 
longitudes' have been determined through a Fourier analysis of the
lightcurves and clearly show migration over several epochs.  The fixed
longitudes found in this work would indicate that is unlikely that
migration could be reproduced by random distributions of spots.  

Another approach for the reconstruction of a surface brightness
distribution is the `two-spot' model.  The `two-spot' model can
account for the light variations of chromospherically active stars and
examples of this model applied to SV Cam can be found in:
\citet{zboril03},\citet{albayrak01}, \citet{kjurkchieva00},
\citet{djurasevic98}, \citet{patkos94}, \citet{budding87} and
\citet{zelik88}.  However Doppler images of chromospherically 
active RS CVn stars (e.g. \citet{petit02hr1099}) show more complicated
distributions of spots than can be reconstructed with the `two-spot'
model. To investigate the reliability of the `two-spot' model
\citet{eaton96} modelled up to 40 randomly distributed spots on 
a single star, and found that they could reproduce `two-spot'
solutions similar to those reconstructed for actual stars.  We have
extended the work of \citet{eaton96} to show that the flux deficit
caused by the presence of many sub resolution spots can be modelled by
two spurious spots at the quadrature points.  

It is concluded that `two-spot' modelling or chi-squared minimisation
techniques are more susceptible to spurious structures being generated
by systematic errors, arising from incorrect assumptions about
photospheric surface brightness, than simple Fourier analysis of the
light-curves.

\section*{Acknowledgments}

The author would like to thank A. Collier Cameron, and V. Holzwarth for
useful discussions, and acknowledges support from a PPARC research
studentship and a scholarship from the University of St Andrews.

\bibliographystyle{mn2e}
\bibliography{iau_journals,master,ownrefs}

\bsp
\label{lastpage}

\end{document}